\documentclass[11pt]{article} 
\usepackage{hyperref} 
\pdfoutput=1
\begin{document} 
\title{The Hungry Fly: Hydrodynamics of feeding in the common house fly}

\author{
        Manu Prakash (manup@stanford.edu)\\
                Junior Fellow, Harvard Society of Fellows\\
        Future home: Department of Bioengineering, Stanford University
            \and
        Miles Steele\\
        Deerfield Academy High School
}


\maketitle

\begin{abstract}
A large number of insect species feed primarily on a fluid diet. To do so, they must overcome the numerous challenges that arise
in the design of high-efficiency, miniature pumps. Although the morphology of insect feeding structures has been described for decades, their dynamics remain largely unknown even in the most well studied species (e.g. fruit fly). Here, in the fluid dynamics video, we demonstrate in-vivo imaging and microsurgery to elucidate the design principles of feeding structures of the common house fly. Using high-resolution X-ray absorption  microscopy, we record in-vivo flow of sucrose solutions through the body over many hours during fly feeding. Borrowing from microsurgery techniques common in neurophysiology, we are able to perturb the pump to a stall position and thus evaluate function under load conditions. Furthermore, fluid viscosity-dependent feedback is observed for optimal pump performance. As the gut of the fly starts to fill up, feedback from the stretch receptors in the cuticle dictates the effective flow rate. Finally, via comparative analysis between the house
fly, blow fly, fruit fly and bumble bees, we highlight the common design principles and the role of interfacial phenomena in feeding.
\end{abstract}


\section{Video Introduction}


The video depicts our experimental work in elucidating feeding mechanisms in insects. The first part of the video depicts feeding mechanics in common house fly {\it (Musca domestica)} using standard video microscopy. Several peculiar observations are made, including extensive gut movement and labirum pulsations. Using a fluid droplet (0.1M sucrose in water) as a lens, we are able to image the entrance point of the labirum; usually closed by filtering flaps. Though static morphology of the insect feeding apparatus is well known - the dynamics of how dilator muscles generate negative pressure is largely unknown. We employ adsorption X-ray microscopy for in-vivo imaging of internal dynamics of several insect species including fruit flies, common house flies and bumblebees - to compare the dynamics of feeding and dependence on various control parameters. X-ray imaging protocols were developed to achieve good contrast in adsorption mode. This allows for low-end X-ray sources to be used for such a task. Second part of the video employs X-ray tomograpghy to accurately measure the 3D morphology of the feeding apparatus (and the entire head) with a imaging resolution of ~10 microns. The third section depicts in-vivo feeding of live common house fly {\it (Musca domestica)} over long durations of time. The pulsatile nature of the cabiral pump is apparent in live video microscopy. We employ micro-surgery techniques from neuro-physiology literature to build probes to perturb the function of the pump. This is done via pulled glass capillary inserted from the antennal plate and the flies are stabilized with PBS solution. Experiments can be conducted for a period of 3 hrs with this adaptation. Long term feeding rate and its dependence to viscosity is also measured. Finally, we compare the pumping mechanisms in various species including fruit flies, blow flies, house flies and bumblebees. Experimental tools developed for in-vivo imaging of insect feeding will also find applications in other insect physiology problems.

\end{document}